# Chemistry and physics of layered oxychalcogenides containing an anti-cuprate type square lattice

Nicola D. Kelly[a]

[a] Cavendish Laboratory, University of Cambridge, J J Thomson Avenue, Cambridge, CB3 0HE, United Kingdom. Email: ne281@cam.ac.uk

**Abstract**

There has been significant recent interest in layered solid-state materials containing an [$M_2$O] square lattice layer ($M$ = transition metal), particularly because [$M_2$O] is the anti-type of the [$CuO_2$] planes in the layered cuprate superconductors. In addition to the superconducting titanium oxypnictides, the [$M_2$O] anti-cuprate layer also occurs in a wide range of layered oxychalcogenide compounds with $M$ spanning early (Ti, V) to later transition metals (Mn, Co, Fe). The chalcogenide in question – which sandwiches the anti-cuprate layer – may be S, Se or Te, and in combination with a wide range of intervening "spacer" layers, many different structural families have been investigated. This review surveys the structures and physical properties of all these oxychalcogenide materials and relates these properties to their common anti-cuprate square lattice [$M_2$O] layer. It is organised around the different oxidation states of the metal ion $M$, in order to explore the effects of the electronic configuration of $M$ on the physical properties of each compound as a whole. A key part of the review highlights the use of soft-chemical modifications to alter physical properties of these materials, in the synthesis of novel van der Waals materials and other metastable compounds. Future avenues for these materials in the bulk, few- and single-layer limits are discussed.

# Table of Contents

## 1 Introduction

The field of solid-state physics and chemistry advanced significantly in 1986 with the discovery of high-temperature superconductivity in the layered cuprates [1,2]. The unusual, non-BCS mechanism for superconductivity is believed to stem from the square-lattice copper oxide [$CuO_2$] planes within the crystal structure, Figure 1(a), with the intervening metal oxide layers acting as the hole- or electron-dopants. In 2012, low-temperature superconductivity was discovered in the oxypnictide material $BaTi_2Sb_2O$ ($T_C$ = 1.2 K) [3]. Researchers noted that the titanium oxypnictide layers in $BaTi_2Sb_2O$ contained a square planar [$Ti_2O$] network, which was the anti-type of the [$CuO_2$] planes in the layered cuprates, i.e. with the cation and anion positions reversed, Figure 1(b). Indeed the crystal structure of the related material, $Na_2Ti_2Sb_2O$ [4], can be described as the exact anti-type of the $La_2CuO_4$ structure: Sb ≡ La, O ≡ Cu, Na ≡ O, Ti ≡ O. Subsequent research efforts have uncovered two-dome superconducting behaviour in the solid solutions $BaTi_2(As_{1-x}Sb_x)_2O$ and $BaTi_2(Sb_{1-y}Bi_y)_2O$ [5], enhancement of $T_C$ with applied pressure [6], and changes to $T_C$ upon isovalent [7] or aliovalent [8–11] chemical doping with or without applied pressure.

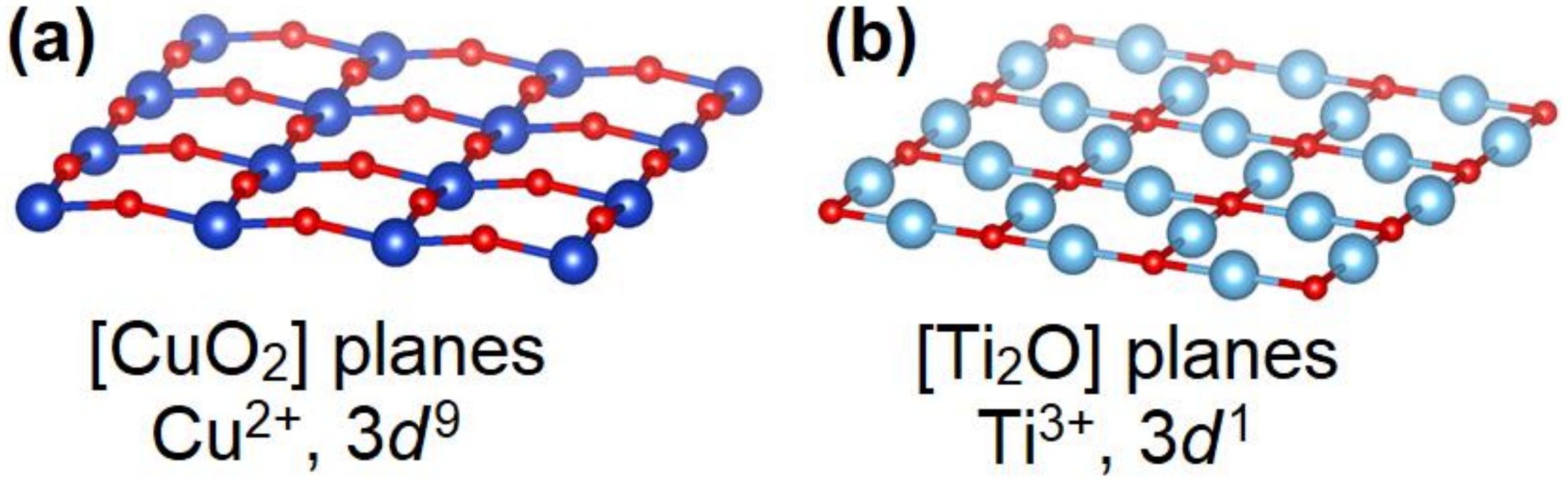

*Figure 1: Arrangement of atoms in (a) $CuO_2$ planes in the cuprate superconductors; (b) $Ti_2O$ planes in the titanium oxypnictide superconductors. Oxide ions are red and metal ions are blue.*

Oxypnictide materials, as an example of mixed-anion materials, are of interest for both pure science and applied research. The combination of two anions with different charge, size, polarisability and chemical bonding preferences typically leads to layered crystal structures, with segregation of the metal cations into "hard" and "soft" layers [12,13]. This enables the development of novel materials for specific applications, e.g. photocatalytic water splitting. In this example, the combination of two anions can produce the "Goldilocks" bandgap, neither too large nor too small, as is the case with typical single-anion materials. Besides oxypnictides, there are a growing a number of oxychalcogenide compounds which show promise in water splitting, catalysis, and more [14,15]. Upon descending Group 16, the ions increase in size and polarisability and decrease in electronegativity (Figure 2). Therefore, in oxychalcogenides containing more than one metal element, the bonding preferences of oxide ions (hard) versus the heavier chalcogenide ions (soft) often create layered structures. For example, in $Sr_2MnO_2Cu_{2-\delta}S_2$ the soft Cu ions preferentially bond with S and the harder Mn ions occupy the oxide layers [16].

| Ion | Radius | $\chi$ |
|---|---|---|
| $O^{2-}$ | 140 pm | 3.44 |
| $S^{2-}$ | 184 pm | 2.58 |
| $Se^{2-}$ | 198 pm | 2.55 |
| $Te^{2-}$ | 221 pm | 2.10 |

*Figure 2: Comparison of the ionic radii (6-coordinate)* [17] *and electronegativities* [18] *of the chalcogenide anions.*

In oxychalcogenide compounds containing the [$M_2$O] anti-cuprate layers, the different charges on S, Se and Te (2–) compared with As, Sb and Bi (3–) in oxypnictides enable a range of different "spacer" layers and a greater structural diversity [19]. The oxypnictides with their highly charged $Pn^{3-}$ ions typically require cations of oxidation state +3 to stabilise the $[M_2Pn_2O]^{2-}$ slab. However, substitution of other cations in the structure (e.g. $Ba_{1-x}Na_xTi_2Sb_2O$) may compensate the charge of $Pn^{3-}$ to some extent [8–11]. The [$M_2Pn_2$O] slab consists of the [$M_2$O] square layer sandwiched between two layers of $Pn^{3-}$ anions (see Figure 3) [20]. Currently, such oxypnictides have been reported with $M$ = Ti, with the few exceptions thus far being the mixed-valent chromium compound $Sr_2Cr_3As_2O_3 \equiv [Sr_2CrO_2]^{2+}[Cr_2As_2O]^{2-}$ [21] and the oxyhydride $La_2Cr_2As_2O_yH_x$ [22]. In contrast, the lower valency of the chalcogenide ions ($Q^{2-}$) allows the inclusion of later transition metals in an equivalent $[M_2Q_2O]^{2-}$ layer. The later transition metals have larger nuclear charge and ionisation energy than early transition metals and therefore tend to favour the +2 oxidation state, especially in solid materials, although examples of $Co^{3+}$, $Fe^{3+}$ and even $Ni^{3+}$ oxychalcogenides are known [23]. Early transition metals are not excluded from the anti-cuprate-type oxychalcogenides, however: in combination with monovalent Group 1 cations, a wide range of Ti- and V-based oxychalcogenides, such as $KV_2Se_2O$, have been synthesised with oxidation states between +2 and +3 [24]. The variety in oxidation states and electron counts across this large family of compounds allows their physical properties to be tuned with composition, e.g. from metal to semiconductor to insulator [25], and novel physics such as altermagnetism has also been explored for these compounds [26,27].

This review systematically discusses the oxychalcogenide compounds containing the anti-cuprate layer [$M_2Q_2$O], where $Q$ is S, Se or Te, and $M$ is a 3$d$ transition metal. First, the structures and physical properties of compounds where $M$ = Mn, Fe and Co are described. These materials all contain $M^{2+}$ ions and are electrically insulating or semiconducting. Open research questions and gaps in the literature are mentioned. Secondly, the focus moves to compounds of the earlier transition metals, Ti and V, which are typically metallic materials containing transition metals of intermediate valency between +2 and +3. This is followed by a section describing how recent advances in soft-chemical techniques have been used to synthesise layered van der Waals compounds and other metastable phases which cannot be accessed using traditional high-temperature methods, and including a discussion of the possibilities for 2D (monolayer) materials derived from these van der Waals compounds. The final section offers a more in-depth discussion of the current gaps in the literature and the compositions of potential new materials. It is hoped that this review will bring together research in both chemical synthetic methods and condensed matter physics experiment and theory, in order to drive forwards the search for new materials.

## 2 Insulating and semiconducting compounds of later transition metals

### *2.1 Structures of the building blocks and overview of known compounds*

This section focuses on the materials containing the mid- to late-transition metal ions Mn, Fe and Co. These ions are denoted $M$ in the general formula [$M_2Q_2$O]$^{2-}$, where $Q$ is a chalcogenide (here limited to S and Se). The [$M_2Q_2$O]$^{2-}$ layer, shown in Figure 3, is built up of edge-sharing *trans*-{$M$O$_2Q_4$} octahedra, and can also be described as an anti-cuprate [$M_2$O] square lattice layer sandwiched between two layers of chalcogenide ions. In these compounds $M$ is always found in the +2 oxidation state.

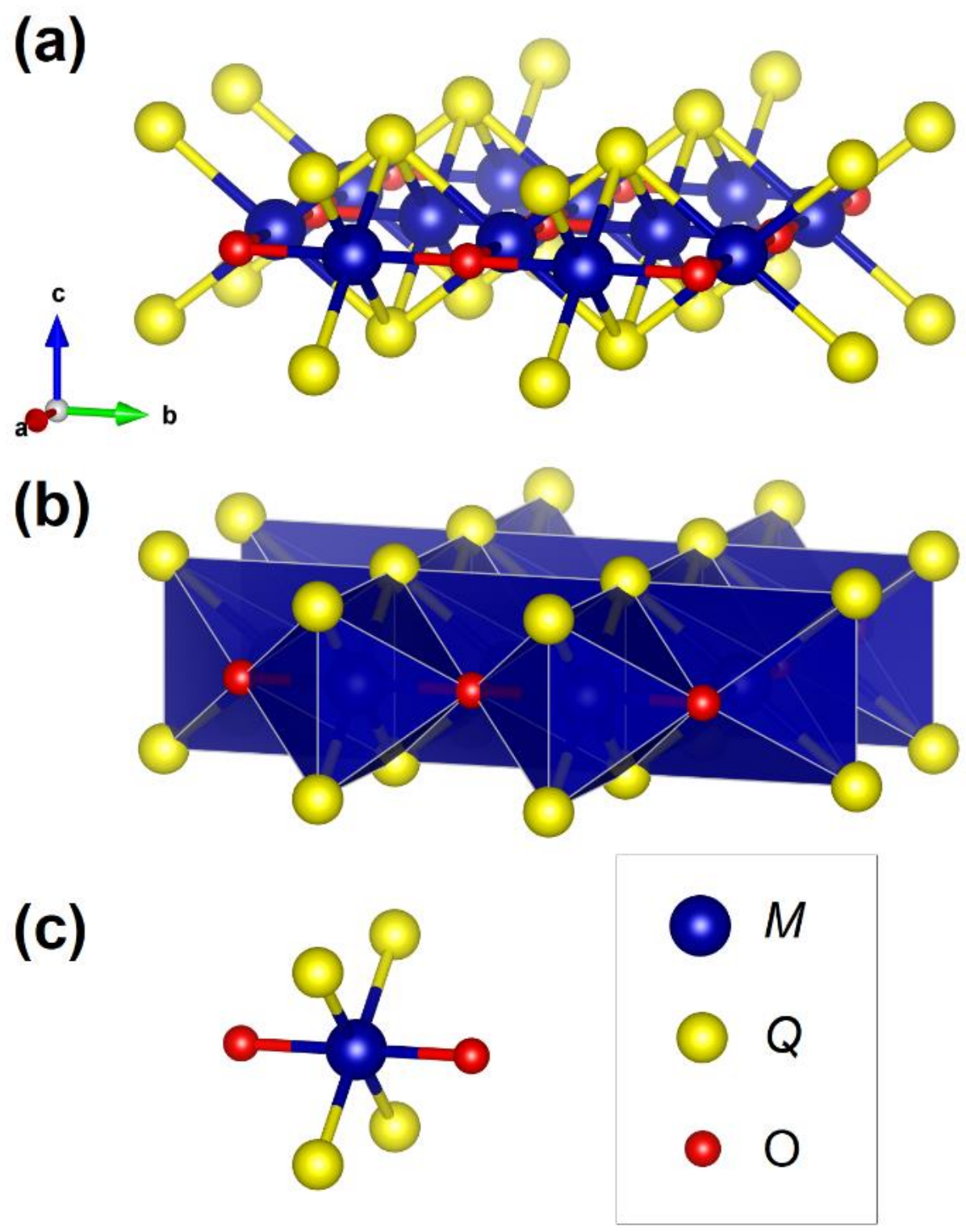

*Figure 3: (a) Crystal structure model of the [$M_2Q_2O$] anti-cuprate layer. (b) Polyhedral view. (c) Coordination environment of each metal ion in an axially distorted octahedron. M ions (3d transition metals) are blue spheres, Q anions (chalcogenides) are yellow and oxygen anions are red.*

For charge balancing, there are several possible intervening "spacer" layers, with a charge of +2, which can combine with the [$M_2Q_2O$]$^{2-}$ layer to build up a neutral compound. Most of the known compounds so far contain spacer layers consisting of $Ba^{2+}$ ions in a single layer, [$Na_2$]$^{2+}$ "double layers", or the fluorite-type blocks [$Ln_2O_2$]$^{2+}$, [$Ba_2F_2$]$^{2+}$ or [$Sr_2F_2$]$^{2+}$, Figure 4.

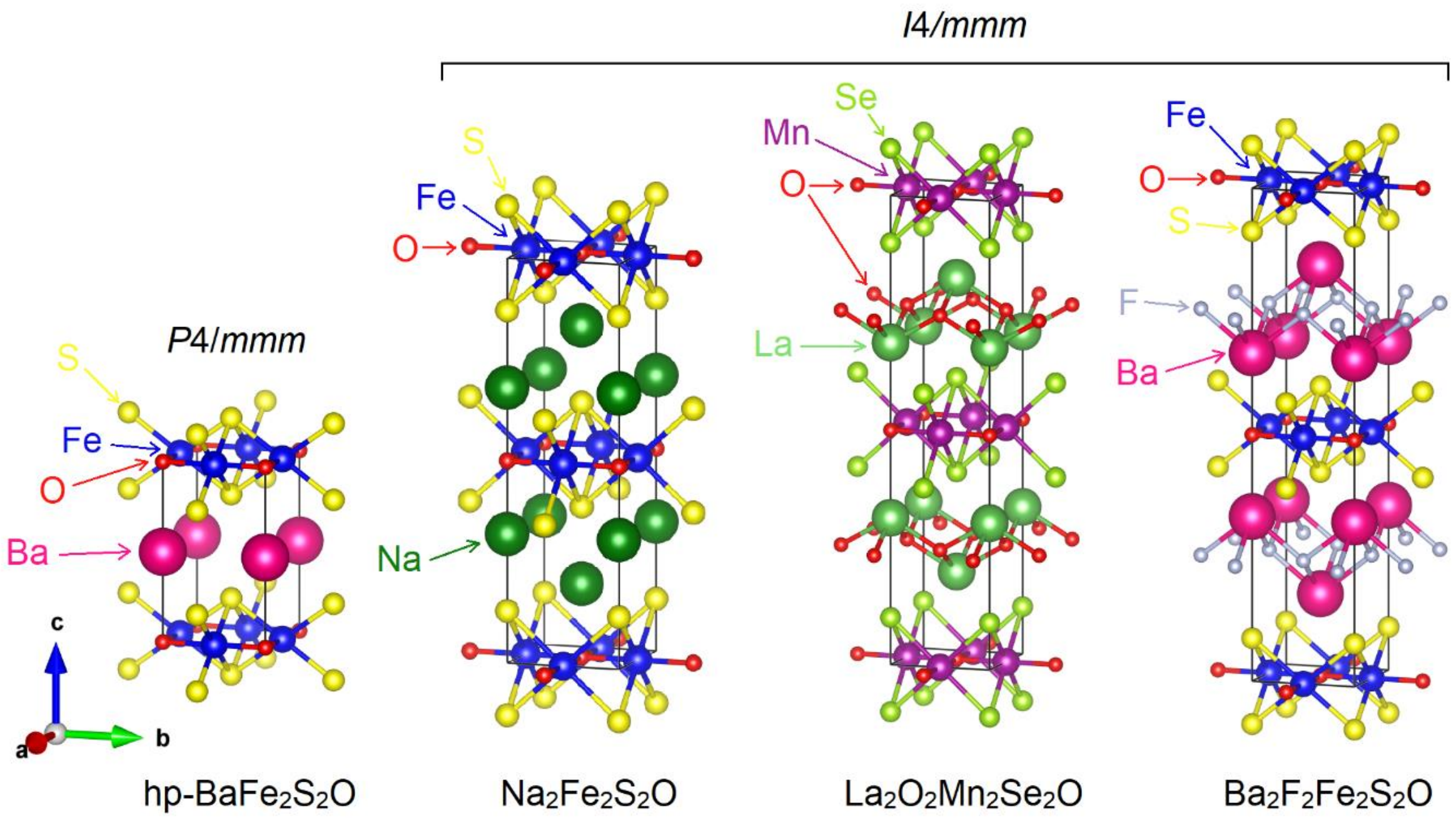

*Figure 4: Crystal structures of some layered oxides containing the anti-cuprate [$M_2Q_2O$] layer, where M is a transition metal cation and Q is a chalcogenide anion. In the labels, "hp" refers to the high-pressure form of $BaFe_2S_2O$.*

The "building block" approach to designing new inorganic materials has been described by Cario and colleagues [19,28] and a similar "modular design" approach was recently employed by Gibson *et al*. to discover another heteroanionic compound $Bi_4O_4SeCl_2$ [29]. In this approach, it is crucial to select building blocks with appropriate, complementary in-plane lattice parameters, charges, and chemical behaviour (bonding preferences) in order to produce a stable layered compound. The fluorite-type $[Ba_2F_2]^{2+}$ or $[Sr_2F_2]^{2+}$ layers are highly flexible and can relax their bond angles in an accordion-like manner in order to match their lattice parameters with other building blocks [19]. Therefore, a range of different metal ions with different ionic radii can be included in the anti-cuprate layer (Table 1 and Figure 5). The same is true of the $[Ln_2O_2]^{2+}$ layers, which can also exhibit a wide range of in-plane lattice parameters through the variation in size of the chemically similar *Ln* ions. In contrast to these or the double-layer $[Na_2]^{2+}$ blocks, it appears that inclusion of the $Ba^{2+}$ ion as a "spacer layer" requires high-pressure synthesis to stabilise the tetragonal $BaFe_2Q_2O$ structure: a different, non-layered structure is observed under ambient-pressure synthesis [30]. These compounds are discussed in section 2.3.

*Table 1: Unit cell parameters of insulating or semiconducting compounds containing the anti-cuprate $M_2Q_2O$ layer: M = Mn, Fe, Co; Q = S, Se. The compounds are grouped by the type of spacer layer, and within each group they are arranged from smallest to largest a-parameter.*

| **Spacer layer type** | **Composition** | **Space group** | **a (Å)** | **c (Å)** | **Reference** |
|---|---|---|---|---|---|
| ▲ Fluorite $[Ba_2F_2]^{2+}$ or $[Sr_2F_2]^{2+}$ *Section 2.2.1* | $Sr_2F_2Fe_2OS_2$ | *I*4/*mmm* | 4.0400(6) | 17.998(4) | [31] |
| | $Sr_2F_2Fe_2OSe_2$ | | 4.0925(2) | 18.5801(10) | [31] |
| | $Ba_2F_2Fe_2OS_2$ | | 4.1238(2) | 19.0885(12) | [31] |
| | $Sr_2F_2Mn_2OSe_2$ | | 4.1658(8) | 18.7310(4) | [32] |
| | $Ba_2F_2Fe_2OSe_2$ | | 4.1946(6) | 19.522(4) | [31] |
| | $Ba_2F_2Mn_2OSe_2$ | | 4.2756(1) | 19.5919(4) | [33] |
| ▼ Fluorite $[Ln_2O_2]^{2+}$ *Section 2.2.2* | $Sm_2O_2Fe_2OSe_2$ | *I*4/*mmm* | 4.00079(7) | 18.3827(5) | [34] |
| | $Nd_2O_2Fe_2OSe_2$ | | 4.0263(1) | 18.4306(2) | [35] |
| | $La_2O_2Fe_2OS_2$ | | 4.0408(1) | 17.8985(6) | [36] |
| | $Pr_2O_2Fe_2OSe_2$ | | 4.04351(4) | 18.4476(3) | [34] |
| | $Ce_2O_2Fe_2OSe_2$ | | 4.05928(9) | 18.4638(5) | [34] |
| | $La_2O_2Co_2OSe_2$ | | 4.0697(1) | 18.4198(2) | [37] |
| | $La_2O_2Fe_2OSe_2$ | | 4.0788(2) | 18.648(2) | [36] |
| | $Pr_2O_2Mn_2OSe_2$ | | 4.09739(2) | 18.69481(9) | [34] |
| | $Ce_2O_2Mn_2OSe_2$ | | 4.11304(2) | 18.74100(9) | [34] |
| | $La_2O_2Mn_2OSe_2$ | | 4.138921(4) | 18.84990(3) | [34] |
| ● $[Na_2]^{2+}$ *Section 2.2.3* | $Na_2Fe_2S_2O$ | *I*4/*mmm* | 4.04222(1) | 14.07319(9) | [38] |
| | $Na_2Fe_2Se_2O$ | | 4.107(8) | 14.641(8) | [39] |
| ■ $Ba^{2+}$ *Section 2.3* | hp-$BaFe_2S_2O$ | *P*4/*mmm* | 4.004(2) | 6.778(4) | [40] |
| | hp-$BaFe_2Se_2O$ | | 4.0748(1) | 7.1501(5) | [41] |

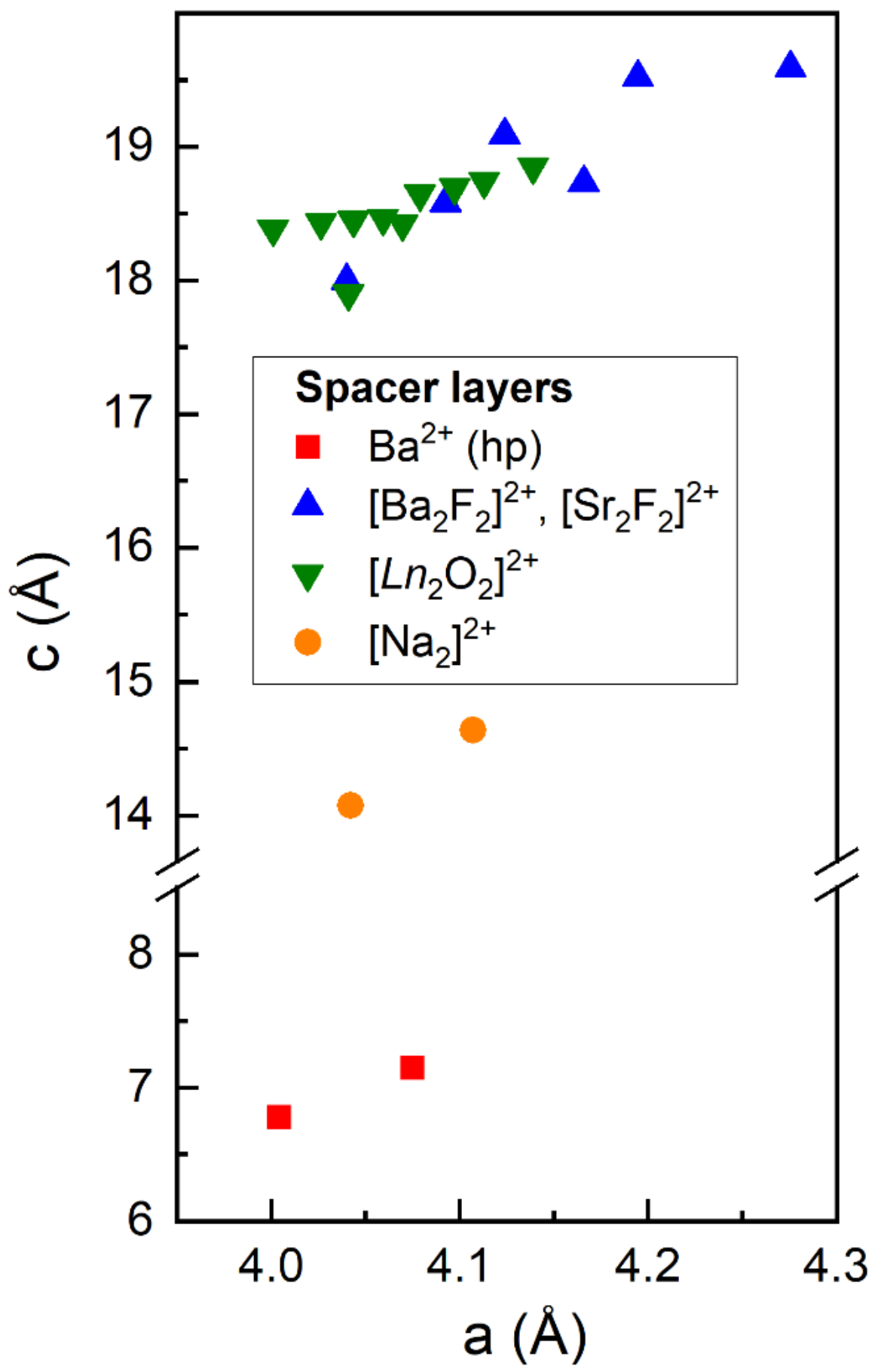

*Figure 5: Unit cell parameters of insulating and semiconducting anti-cuprate compounds reported in the literature. Note that a similar plot for the metallic compounds discussed in section 3.2 may be found in reference* [24].

The following sections describe the synthesis, structures and physical properties of the materials $AM_2Q_2O$ with different $A$ spacer layers.

## 2.2 *Synthesis, structures and physical properties of the layered insulators $AM_2Q_2O$ where $M$ = Mn, Fe, Co; $Q$ = S, Se*

### 2.2.1 *$A$ = [$Ba_2F_2$], [$Sr_2F_2$]*

Among the first anti-cuprate oxychalcogenides to be investigated were the iron-based compounds $Ba_2F_2Fe_2Q_2O$ and $Sr_2F_2Fe_2Q_2O$ ($Q$ = S, Se) in 2008 [31]. These four compounds are isostructural, crystallising in the body-centred tetragonal crystal structure shown in Figure 4. They had been predicted to exist using the "building blocks" approach [19,28], combining the $[Fe_2Q_2O]^{2-}$ anti-cuprate layer with a fluorite-like $[Ba_2F_2]^{2+}$ layer. Successful experimental synthesis and characterisation followed, revealing that the four compounds are antiferromagnets with Néel temperatures ($T_N$) between 80 and 110 K. A clear trend in $T_N$ with lattice parameters was observed: as the in-plane lattice parameter $a$ decreases (due to chemical substitution), the $T_N$ increases, representing stronger spin-spin interactions. Resistivity data obtained for $Ba_2F_2Fe_2Se_2O$ and $Sr_2F_2Fe_2S_2O$ indicated Mott insulating behaviour with electronic conductivity arising from hopping of unpaired electrons between nearby $Fe^{2+}$ ions. The conductivity was 2 orders of magnitude larger for $Sr_2F_2Fe_2S_2O$ than $Ba_2F_2Fe_2Se_2O$ and this was also attributed to the shorter hopping distances between Fe ions, which is correlated with the lattice parameter. The specific heat data above and below $T_N$ indicate that the 2D Ising model is appropriate for these materials, because the magnetic specific heat $C_{mag}$ varies linearly with $\log|(1 - T_N)/T_N|$ both above and below the ordering transition. Moreover, the authors measured $^{57}$Fe Mössbauer spectra, extracted the hyperfine magnetic field values $H$, and obtained $\beta$ values of 0.118 ($Ba_2F_2Fe_2Se_2O$) and 0.15 ($Sr_2F_2Fe_2S_2O$), where $\beta$ is the slope of the plot of $\log(H/H_{sat})$ against $\log(1 - (T/T_N))$. These are consistent both with the Ising model (expected $\beta = 0.125$) and with the temperature variation of the magnetic moment obtained from neutron diffraction: $M_{Fe} = C(1 - (T/T_N))^{\beta}$, $\beta = 0.15$ ($Ba_2F_2Fe_2Se_2O$). Such Ising behaviour would not be expected for $Fe^{2+}$ ions in a perfectly octahedral coordination environment; however, in the anti-cuprate materials $Fe^{2+}$ is coordinated to 2 x $O^{2-}$ and 4 x $Q^{2-}$ in a distorted octahedron, and this strong anisotropy permits the Ising antiferromagnetism.

Kabbour *et al.* also used DFT calculations (GGA+U) to predict possible magnetic ground states of $Ba_2F_2Fe_2Q_2O$ and $Sr_2F_2Fe_2Q_2O$ [31]. There is competition between several different exchange interactions in this system, as shown in Figure 6(a). Different authors have chosen different labels for the interactions, e.g. $J_1/J_2/J_3$ [31] versus $J_1/J_2/J_2$' [42]. In order for $J_1$ to refer to the nearest-neighbour interaction (shortest distance in space), here we follow the notation of Zhu *et al.* [43] and others [34,42] in choosing to use $J_1$ (Fe-O-Fe or Fe-S-Fe nn superexchange), $J_2$ (Fe-S-Fe nnn superexchange), and $J_2$' (Fe-O-Fe nnn superexchange); nn = nearest neighbours, nnn = next-nearest neighbours. The DFT calculations predict antiferromagnetic (AFM) interactions $J_2$' and $J_1$, whereas the $J_2$ interactions are ferromagnetic (FM). Of these, the dominant interaction is the AFM $J_2$' representing 180° Fe-O-Fe superexchange; this agrees with the Goodenough-Kanamori rules [44]. However, the presence of both $J_1$ and $J_2$' with similar magnitudes leads to geometric frustration: the system can be described as a frustrated checkerboard square lattice, where long-range order occurs – despite the frustration – because of the large spin, $S = 2$, for high-spin $Fe^{2+}$ ions.

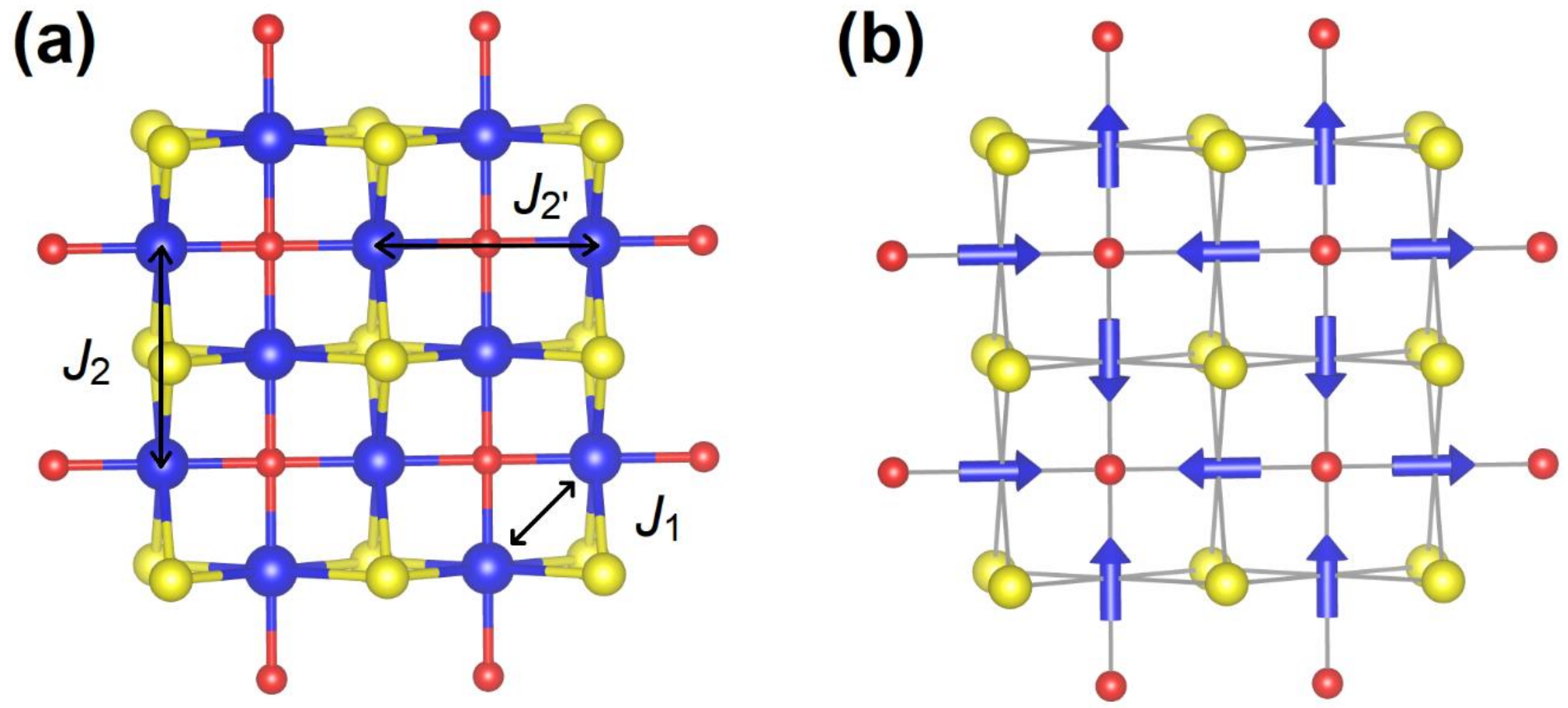

*Figure 6: (a) Exchange interactions $J_1$, $J_2$ and $J_2'$ in the [$Fe_2Q_2O$] anti-cuprate layer. (b) Illustration of the 2-**k** magnetic model observed for several compounds in this family.*

The magnetic structures for materials containing the [$Fe_2Q_2O$] anti-cuprate layer have been explored using neutron diffraction and Mössbauer spectroscopy. Zhao *et al*. investigated $Sr_2F_2Fe_2S_2O$ using neutrons [45] and determined its magnetic structure, which is represented in Figure 6(b). The $Fe^{2+}$ moments display strong Ising anisotropy, pointing along the directions of Fe–O bonds within the anti-cuprate planes; combined with the competing exchange interactions, this yields a complex magnetic structure. The structure can be described in words as orthogonal chains of antiferromagnetically coupled spins, and mathematically by two orthogonal **k**-vectors, (½, 0, ½) and (0, ½, ½).[1] An alternative single-**k** model with all spins collinear was considered but discounted because, despite producing an equally good fit to the magnetic diffraction data, the ordered moment on each of the two Fe sites in that model was too small. In contrast, the 2-**k** model gave a larger ordered moment (by a factor of $\sqrt{2}$) which was more consistent with the spin-only prediction. Furthermore, previous Mössbauer spectroscopy data had indicated a single $Fe^{2+}$ site [31]. The 2-**k** model is consistent with the Goodenough-Kanamori rules, in that the $J_{2'}$ interactions (180° via oxide ions) are antiferromagnetic while the $J_2$ interactions (~90° via sulfide ions) are ferromagnetic, and with the DFT calculations mentioned earlier [31].

Two manganese compounds with the [$Ba_2F_2$] and [$Sr_2F_2$] spacer layers were reported in 2011 and 2013 respectively. $Ba_2F_2Mn_2Se_2O$ is a semiconductor and an antiferromagnet with $T_N$ = 100 K; a broader maximum in the magnetic susceptibility was observed at 210 K [33].

[1] On moving to subsequent layers of Fe ions (along the *c*-axis), the spins are rotated in the *ab* plane by 90°. Thus there are four Fe layers in the magnetic unit cell, which is twice as tall as the nuclear cell.

$Sr_2F_2Mn_2Se_2O$ is isostructural with its Ba analogue but has a much smaller bandgap and lower resistivity due to the substitution of less electropositive Sr atoms [32]. Both compounds are believed to develop long-range order at $T_N$ (100 K and 141 K respectively) from a low-dimensional or short-range AFM ordered state at $T > T_N$, arising from strong geometric magnetic frustration. A considerable amount of entropy is associated with the short-range ordering, such that the peaks in heat capacity at $T_N$ are rather small: much smaller than those observed for related Fe and Co compounds, suggesting that the Mn compounds experience greater frustration than the Fe and Co analogues. No neutron diffraction data have yet been reported for $Ba_2F_2Mn_2Se_2O$ or $Sr_2F_2Mn_2Se_2O$, but the related compound $La_2O_2Mn_2Se_2O$ (see below, section 2.2.2) displays G-type AFM order as well as 2D short-range correlations at higher temperatures [46]. It therefore appears, from the data collated in this and the following sections, that the 2-**k** magnetic model is unique to the Fe-containing anti-cuprate compounds.

#### *2.2.2 A = [$Ln_2O_2$]*

Isostructural with the [$Ba_2F_2$] and [$Sr_2F_2$] spacer layers is another fluorite-type layer, [$Ln_2O_2$], which also has a net +2 charge. Two materials containing [$Ln_2O_2$] layers had been reported by Mayer *et al.* in 1992: $La_2O_2Fe_2S_2O$ and $La_2O_2Fe_2Se_2O$, which were found to be semiconductors with signatures of 2D magnetic order around 100 K [36]. Later studies by Fuwa *et al.* [35] and McCabe *et al.* [42] determined that $La_2O_2Fe_2Se_2O$ displays the 2-**k** (non-collinear) magnetic model, as described in the previous section (2.2.1, Figure 6(b)). This model, with its 2D Ising behaviour, was confirmed by Günther *et al.* using local probes, $^{139}La$ nuclear magnetic resonance (NMR) and muon-spin relaxation spectroscopy, giving a critical exponent $\beta = 0.133$ [47]. Analogues with later lanthanide ions have also been produced including *Ln* = Ce, Pr, Nd, and Sm [34,35]. The 2-**k** magnetic structure has been reported to fit experimental data (neutron diffraction and/or Mössbauer spectroscopy) for $Ce_2O_2Fe_2Se_2O$, $Nd_2O_2Fe_2Se_2O$, and the oxysulfide analogue, $La_2O_2Fe_2S_2O$ [48]. In their report on $La_2O_2Fe_2S_2O$, Oogarah *et al.* commented on the robustness of the 2-**k** model across several different iron oxychalcogenide materials with different lattice parameters, including both sulfide and selenide materials. The model is dominated by the Ising anisotropy of the $Fe^{2+}$ moments and is much less frustrated than the alternative single-**k** model, since both the strong next-nearest-neighbour $J_2$ (FM) and $J_2'$ (AFM) interactions are satisfied. The onset of AFM order upon cooling is typically preceded by 2D short-range order (as observed by a Warren-type peak in neutron diffraction data) for 10 to 20 K above $T_N$ [48]. The 2-**k** model appears again in the high-pressure polymorphs of $BaFe_2S_2O$ and $BaFe_2Se_2O$, Section 2.3. These results further demonstrate that the 2-**k** magnetic structure is an intrinsic feature of the [$Fe_2Q_2O$] anti-cuprate layers, regardless of the intervening building blocks.

Free *et al*. [34] synthesised a series of manganese-containing compounds $Ln_2O_2Mn_2Se_2O$ (*Ln* = La, Ce, Pr) with the same crystal structure as the iron-containing compounds. Using variable temperature powder X-ray and neutron diffraction, they found a structural distortion upon cooling: the rate of thermal expansion in the *c*-direction changed sharply (at 115, 90, and 180 K for La, Ce, Pr compounds respectively) and there was an increase in the anisotropic thermal displacement parameter $U_{33}$ (i.e., the *c*-direction) for the oxide ions within the anti-cuprate layers. This was modelled using a split site for those oxide ions, with a maximum displacement of around 0.3 Å out of the plane at the lowest temperature of 12 K, which indicates that the Mn–O–Mn bond angles are lowered from their ideal value of 180° to around 163°. Interestingly, the onset temperature and magnitude of this structural distortion appeared not to be correlated with the size of the lanthanide ion. Small superstructure peaks appeared in both X-ray and neutron data at low temperatures, which were attributed to ordering of these oxide ions. A similar local distortion, i.e. buckling of the [$M_2O$] plane as the oxide anions are displaced, has been found in the iron compounds $La_2O_2Fe_2Se_2O$ and $La_2O_2Fe_2S_2O$ by Karki *et al*. [49]. Their analysis of the neutron pair distribution functions (PDFs) also revealed largely temperature-independent, short-range orthorhombic distortions on the order of 10–20 Å in both $La_2O_2Fe_2Se_2O$ and $La_2O_2Fe_2S_2O$, with no significant difference in orthorhombicity between the selenide and sulfide materials. The occurrence of similar structural distortions in both Fe- and Mn-containing materials suggests that this behaviour may be widespread in the anti-cuprate family. Now that we know to look for it, future studies should examine the oxygen ion positions and thermal parameters more closely, using either neutron diffraction (fitting the average structure) or total scattering/PDF analysis (fitting the local structure), in other materials in this family.

The magnetic properties of the $Ln_2O_2Mn_2Se_2O$ series are quite different from their Fe analogues. $La_2O_2Mn_2Se_2O$ displays commensurate AFM order below 168 K, with the $Mn^{2+}$ spins directed parallel to *c* in a checkerboard arrangement. $Ce_2O_2Mn_2Se_2O$ and $Pr_2O_2Mn_2Se_2O$ have the same magnetic structure with $T_N$ = 174 and 180 K respectively, where the trend in transition temperatures likely reflects an increasing orbital overlap with the decreasing ionic radius of $Ln^{3+}$. This magnetic structure is dominated by the antiferromagnetic $J_1$ interaction (~90° superexchange pathway via oxide ions) [34]. $La_2O_2Mn_2Se_2O$ has also been investigated by Ni *et al*. in the bulk [46] and Liu *et al*. using single crystals [33]; both groups reported physical properties consistent with those of Free *et al.*, i.e., a long-range AFM transition on cooling, preceded by short-range ordering. Furthermore, $Pr_2O_2Mn_2Se_2O$ undergoes symmetry lowering from tetragonal (*I*4/*mmm*) to orthorhombic (*Immm*) below 36 K [34]. This structural transition is believed to be driven by crystal electric field effects of the non-Kramers $Pr^{3+}$ ion ($4f^2$ electron configuration) since it has also been observed in $Pr_2O_2Fe_2Se_2O$ [50] and two other layered compounds, PrMnSbO [51] and $PrMnAsO_{1-x}F_x$ [52], at similar temperatures. However, $Pr_2O_2Mn_2Se_2O$ and $Pr_2O_2Fe_2Se_2O$ showed no evidence of long-range order of $Pr^{3+}$ moments at low temperatures in zero-field neutron diffraction data [34,50] and in general, the $(Ln_2O_2)M_2Q_2O$ compounds discussed in this section are dominated by the transition-metal rather than the rare-earth magnetism.

More recently, $La_2O_2Mn_2Se_2O$ was proposed to be a correlated insulating d-wave altermagnet. Altermagnets are a recently defined class of magnetic materials with alternating spins on adjacent sites, and zero net magnetisation, like an antiferromagnet. However, the up- and down-spin sublattices are connected by a real-space rotation transformation, rather than a translation or inversion as in classical antiferromagnets. This lifts the Kramers spin degeneracy and leads to unusual magnetic properties [53]. Wei *et al.* identified that the checkerboard pattern of $Mn^{2+}$ spins in $La_2O_2Mn_2Se_2O$ can be described using two sublattices which, because of the positions of oxide and selenide anions, are not related by translational or inversional symmetry. This makes the material an altermagnet, whereas the Fe (2-**k** model, see above) and Co analogues (see below) retain translational symmetry and are not altermagnetic [54]. They also used high-temperature Curie-Weiss fitting to confirm the high-spin configuration of $Mn^{2+}$, and neutron PDF measurements to analyse the short-range ordering above $T_N$. The recent studies into $La_2O_2Mn_2Se_2O$ and $KV_2Se_2O$ (section 3.2) suggest that the family of anti-cuprate layered oxychalcogenides may be a rich source of altermagnetic materials. This is sure to be a rapidly growing research area in the coming years.

Finally, $La_2O_2Co_2Se_2O$ has also been synthesised and characterised, almost simultaneously by three independent groups of researchers in 2010–11. Wang *et al.* observed long-range AFM order at 220 K with a broad feature in the susceptibility around 250 K, attributed to short-range ordering. They fitted the data at 320–400 K to the Curie-Weiss formula and obtained $\mu_{eff} = 1.68$ $\mu_B$ per Co, which would correspond to low-spin $Co^{2+}$ ($S = 1/2$) [55]. However, Fuwa *et al.* fitted their own susceptibility data (which had the same form) to the 2D Heisenberg model and obtained a good fit by assuming high-spin $Co^{2+}$ ($S = 3/2$) [37]. Free *et al.* then carried out a neutron diffraction study which confirmed that the $Co^{2+}$ ions are high-spin (refined moment:

3.29(3) $\mu_B$ per Co) [34]. It is likely that the paramagnetic Curie-Weiss law does not hold in the temperature range used by Wang *et al*. because of the high $T_N$, such that there are still persistent short-range correlations around room temperature.

The proposed magnetic structure of $La_2O_2Co_2Se_2O$ at low temperatures is in some ways similar to the 2-**k** model for the Fe compounds, in that the spins all lie in the *ab* plane and the $J_1$ nearest-neighbour pairs of spins are all orthogonal [56]. However, the model for the Co compound has both the $J_2$ and $J_2$' interactions AFM, with the magnetic moments directed either parallel or perpendicular to the Co–O bonds (always within the anti-cuprate planes). The two models cannot be distinguished using powder data [34]. Both models are different from either the Fe 2-**k** model (where $J_2$ is FM) or the Mn compounds where the spins are parallel to *c*, demonstrating the great variety in magnetic behaviour that can be produced in the anti-cuprate family by varying the transition metal cation. Future work in synthesising new Co and Mn oxychalcogenides, by changing the intervening "spacer" layers, and elucidating their magnetic structures may provide further examples of these different models and possibly evidence for global, transition metal-dependent magnetic structures of the anti-cuprate layer.

### 2.2.3 *A = [Na₂]*

There are two known oxychalcogenides which contain a double layer of sodium ions between the iron-based anti-cuprate layers. $Na_2Fe_2Se_2O$ was targeted as an oxychalcogenide analogue of the titanium oxypnictides such as $Na_2Ti_2Sb_2O$, and was successfully synthesised in 2011 by He *et al*. [39]. It crystallises in space group *I*4/*mmm* with ABAB stacking of the anti-cuprate planes (Figure 4). It is an insulator, and its magnetic behaviour is similar to many of the other iron oxychalcogenides mentioned in earlier sections, i.e. a broad hump in the magnetic susceptibility and a more subtle feature at a lower temperature, $T_N$ = 73 K, which were assigned to short-range and long-range AFM ordering respectively. The short-range correlations, which are common in layered materials, persist to a temperature approximately twice $T_N$. Similar magnetic properties were observed for $Na_2Fe_2S_2O$ which has a slightly higher $T_N$ = 91.5 K; this was suggested to stem from the shorter Fe–S distance compared with Fe–Se [57]. Both $Na_2Fe_2S_2O$ and $Na_2Fe_2Se_2O$ have been investigated as redox cathodes for rechargeable batteries owing to the high earth abundance of Na and Fe [38,58]. To date, the magnetic structures of these materials at low temperatures have not been investigated. It should be possible to determine these using neutron diffraction and/or Mössbauer spectroscopy.

### *2.3 High-pressure synthesis of $BaFe_2Q_2O$*

Traditional ceramic synthesis, from BaO, Fe, and S or Se in an evacuated quartz ampoule, yields orthorhombic crystals with the composition $BaFe_2S_2O$ or $BaFe_2Se_2O$ respectively. These materials are Mott insulators and exhibit a fascinating spin-ladder magnetic structure [30,59–61]. Upon application of high pressure and temperature (3.5 GPa, 950 °C, 2 h), orthorhombic $BaFe_2S_2O$ transforms into the tetragonal *P*4/*mmm* phase with anti-cuprate layer structure, as shown in Figure 4, and can be recovered as a stable phase at ambient pressure after the high-pressure synthesis. Interestingly, the analogous oxyselenide hp-$BaFe_2Se_2O$ was produced directly from BaO, Fe and Se at 1000 C and 5 GPa, i.e. without any pre-reaction in a sealed tube. It may be possible also to synthesise hp-$BaFe_2S_2O$ directly at high pressure, but no such attempts have been reported.

The tetragonal polymorph of $BaFe_2S_2O$ [40] is a Mott insulator. It exhibits Néel AFM order below $T_N$ = 121 K, and calculations suggest that the 2-**k** model discussed previously (section 2.2) is appropriate. It also has a second magnetic transition at around 40 K which is associated with the development of weak ferromagnetism, as seen by the bifurcation of ZFC and FC susceptibility data and small hysteresis in M(H) at low temperatures. The small size of the net moment suggests that this ferromagnetism arises from Dzyaloshinskii-Moriya superexchange interactions and spin canting. In the analogous selenide, tetragonal hp-$BaFe_2Se_2O$, Takeiri *et al.* found a similar 2-**k** magnetic structure below $T_N$ = 106 K and a tiny net moment of 0.002 $\mu_B$/Fe at T = 2 K, again suggesting slight spin canting [41]. No further examples of $BaM_2Q_2O$ with other *M* ions have yet been reported, nor the telluride analogue $BaFe_2Te_2O$.

## 3 Metallic compounds of early transition metals

### *3.1 Overview of known compounds*

This section discusses layered compounds containing early transition metals (Ti, V) in the *M* sites of the [$M_2Q_2O$] anti-cuprate layers. Across each transition metal series of the periodic table, the $n^{th}$ ionisation energy increases as the effective nuclear charge increases, making it more difficult to remove an electron from the atom. Therefore, higher oxidation states can be more easily stabilised by early transition metals.

The compounds discussed in this section have the general formula $AM_2Q_2O$ where *M* is Ti or V; *Q* is a chalcogenide anion (S, Se or Te); and *A* is a heavy alkali metal ion (K, Rb or Cs). Assuming the oxidation states of –2 for O and *Q*, +1 for A, we are left with a total of +5 for [$M_2$], i.e. an average metal oxidation state of +2.5. Importantly, these are not *mixed-valent* but rather *intermediate valency* compounds: rather than a mixture of +2 and +3 ions, there is only a single crystallographic *M* site, so the *M* ions must all be identical and indistinguishable [62]. As such, these materials tend to be black in colour and good conductors of electricity.

Between 2016 and 2021, five compounds in this family were synthesised: $CsV_2S_2O$ [63], $CsTi_2Te_2O$ [64], $CsV_2Se_2O$ [25], $RbV_2Te_2O$ [65] and $KV_2Se_2O$ [66]. In 2023, Kelly and Clarke completed the set of 3 x 2 x 3 = 18 compositions achieved by systematically varying *A*, *M* and *Q* [24]. In general, all of the members of this family behave as metals or bad metals and temperature-independent paramagnets; many are air-sensitive, particularly the tellurides, which easily oxidise to produce elemental tellurium [24,64]. Some notable properties of selected materials are discussed in the following sections.

### 3.2 *Synthesis and chemical and physical properties of the metals (K/Rb/Cs)$M_2Q_2$O where M = Ti, V; Q = S, Se, Te.*

The titanium and vanadium oxychalcogenides are typically synthesised using traditional ceramic methods at high temperatures (700–800 °C). Reactions are carried out in evacuated quartz ampoules to avoid oxidation of the elemental alkali metals, especially Cs which is highly reactive and sometimes reacts with the other reagents while still under an argon atmosphere [24]. Several authors reported alkali metal deficiencies of up to about 20 %, e.g. $Rb_{1-\delta}V_2Te_2O$ [65], $Cs_{1-x}Ti_2Te_2O$ [64], which were difficult to control synthetically. Compared with the stoichiometric compounds, these alkali metal deficiencies lead to an increase in the average oxidation state of the transition metal to maintain charge neutrality, which has two effects. First, the lattice parameter *a* (which corresponds to two *M*–O bonds, as can be seen in Figure 4) decreases as the *M* ionic radius decreases, while the *c* parameter typically increases because some of the strong ionic inter-layer interactions are replaced by weaker van der Waals interactions. This effect can be observed in the reactions of $KV_2Se_2O$ or $KV_2S_2O$ with moist air, causing gradual de-intercalation of the potassium ions, Figure 7. Secondly, the alkali metal deficiency effectively dopes holes onto the anti-cuprate layers, making them closer to the $Ti^{3+}$, $3d^1$ state found in the barium titanium oxypnictides like $BaTi_2Sb_2O$. However, to date no superconductivity has been observed in the titanium or vanadium oxychalcogenides, unlike their oxypnictide relatives [20].

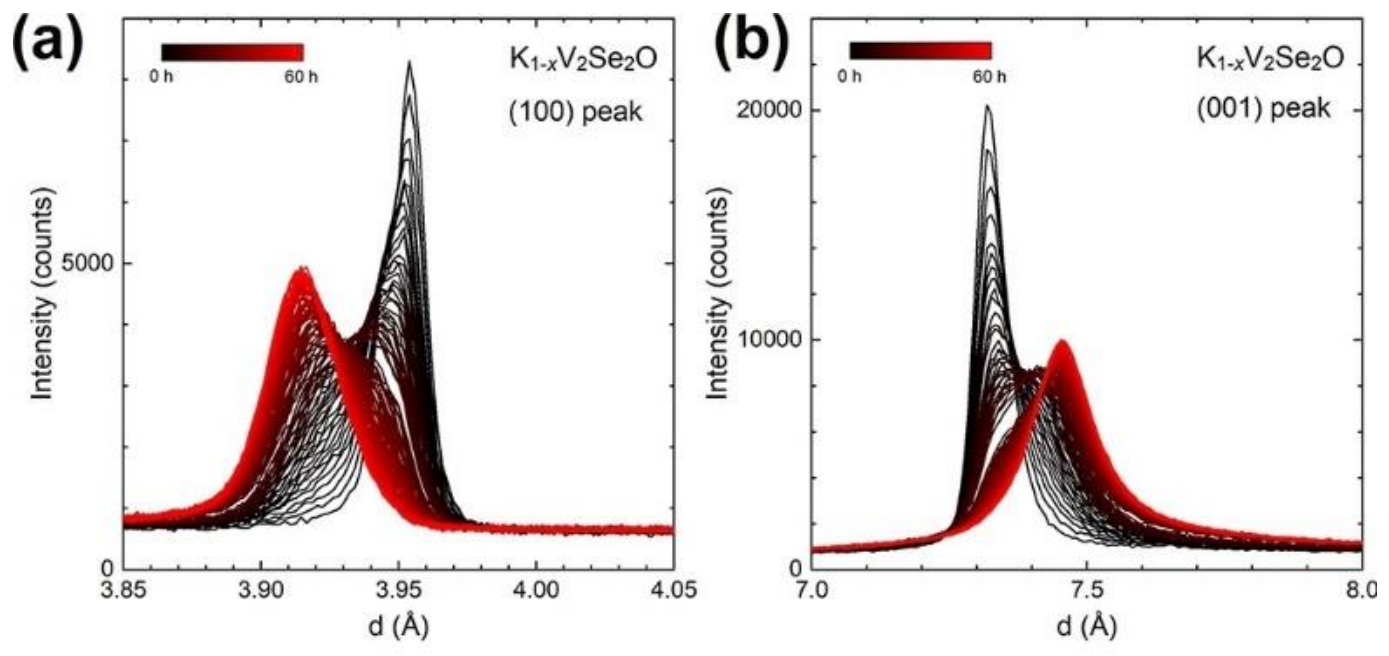

*Figure 7: PXRD patterns for $KV_2Se_2O$ upon air exposure, which causes gradual de-intercalation of $K^+$ ions. As the K occupancy decreases over time, so the a-parameter decreases ((100) peak) and the c-parameter increases ((001) peak). Adapted from reference* [24].

Spin- or charge-density-wave transitions have been observed in several of the titanium and vanadium oxychalcogenides, similar to those found in the titanium oxypnictides. For example, a metal-metal transition was observed in $Rb_{1-\delta}V_2Te_2O$ at 100 K [65], and $CsV_2Se_2O$ has an anomaly at 168 K [25]. Furthermore, Lin *et al*. were able to tune semiconducting $CsV_2Se_2O$ to metallic $CsV_2Se_{1.5}O$ by varying the Se content, which reduces the average oxidation state of vanadium to +2, its lowest stable non-zero value in the solid state [25]. Most recently, research efforts have focused on $KV_2Se_2O$. In 2024 Bai *et al*. reported its detailed physical characterisation including a density-wave-like transition at 105 K in the magnetic susceptibility, heat capacity, and resistivity data, yet no structural distortions were observed in their variable-temperature X-ray data [67]. Further electron diffraction data from Zhuang *et al*. confirmed the absence of superlattice peaks or diffuse scattering, and therefore the absence of a structural transition at the temperatures where the density-wave transition occurred [68]. Their experimental results were complemented by a density functional theory (DFT) study which suggested that, as the unit cell parameters decrease upon cooling, charge transfer occurs from O to V ions which decreases the oxidation state of vanadium. This was proposed as a possible explanation for the anomalies in heat capacity and resistivity [68].

However, in a recent study, Jiang *et al*. also proposed $KV_2Se_2O$ as an altermagnet candidate, like $La_2O_2Mn_2Se_2O$ (section 2.2.2). The key signatures for a metallic altermagnetic state in $KV_2Se_2O$ were described as a spin-polarised band structure and highly anisotropic Fermi surface, with evidence for d-wave exchange splitting at the SDW transition provided by $^{51}V$ NMR [27]. Furthermore, the high anisotropy generates a giant spin current and a polarised electric current, with potential applications in spintronic devices. In a recent preprint, Yan *et al*. used DFT calculations and both angle-resolved photoemission spectroscopy (ARPES) and physical property measurement system (PPMS) experiments to confirm the d-wave altermagnetic behaviour and to explain the symmetry breaking at the SDW transition using a spin canting model [69].

In order to understand the physics of $KV_2Se_2O$ more, and to compare it to substituted analogues such as $KV_2S_2O$ and $KV_2Te_2O$ [24], it is suggested that the resistivities of all the metallic "1221" compounds should be measured in a pressure cell. Unlike using chemical pressure (substitution) to tune the physical properties, the pressure cell is a "clean", closed environment which would enable changes in lattice parameters to be studied without the possibility of $K^+$ de-intercalation or other decomposition, which might alter the vanadium oxidation state. In these compounds where the alkali metal components have been shown to be labile, such experiments will be crucial for understanding the intrinsic material properties and developing the exciting new field of anti-cuprate type altermagnets.

## 4 Soft-chemical methods for transformation of layered oxychalcogenides

### *4.1 Introduction to soft chemistry and overview of oxychalcogenide examples.*

Soft chemistry, or "chimie douce", refers to reactions carried out at ambient pressure and near room temperature, typically using mild reagents [70]. In contrast with the traditional "shake-and-bake" high-temperature ceramic synthesis, which favours the thermodynamic products, soft-chemical reactions can be employed to produce kinetically stabilised phases that might not be easily obtained by ceramic methods. Many such reactions are also *topochemical*, i.e., the intrinsic structures of the layers or other building blocks of the parent material are maintained whilst atoms or molecules are inserted or removed between the layers. This contrasts with high-temperature ceramic methods, where the entire crystal structure of the precursors is destroyed as atoms diffuse across grain boundaries to form the eventual product structure.

This section discusses first the topochemical synthesis of van der Waals phases $M_2Q_2O$ ($M$ = Ti, V; $Q$ = Se, Te) via oxidative deintercalation of alkali metal ions from $AM_2Q_2O$, followed by the reductive intercalation of lithium into $V_2Te_2O$ by both chemical and electrochemical routes. The physical properties of the three van der Waals compounds are compared and contrasted. Finally, recent progress in 2D anti-cuprate materials is discussed.

### *4.2 Synthesis, structures and properties of the van der Waals phases $V_2Se_2O$, $V_2Te_2O$ and $Ti_2Te_2O$.*

Lin *et al.* first reported $V_2Se_2O$ in 2018 [25]. It was synthesised from the parent $CsV_2Se_2O$ (Section 3.2) by the following reaction, at 60 °C for 72 h in an autoclave:

$$CsV_2Se_2O + \text{excess } I_2 \longrightarrow V_2Se_2O + CsI$$

Formation of the stable ionic salt CsI drives the reaction. $Cs^+$ ions are extracted from $CsV_2Se_2O$ and the remaining anti-cuprate layers shift within the *ab* plane, such that the product $V_2Se_2O$ has the body-centred space group *I*4/*mmm* instead of primitive *P*4/*mmm*, Figure 8. The reaction has a significant effect on the physical properties: $CsV_2Se_2O$ is a semiconductor, while $V_2Se_2O$ is a strongly correlated insulator with an unusual $\rho \propto \log(1/T)$ dependence over a wide temperature range [25]. As shown in Figure 8, the crystal structure of $V_2Se_2O$ is built up from anti-cuprate layers stacked in an ABAB manner in the *c*-direction. Adjacent layers are held together by weak van der Waals interactions Se….Se, as in the well-known transition metal diselenides $MSe_2$. This makes it an ideal platform for studying possible exfoliation or intercalation reactions, since the layers are weakly coupled.

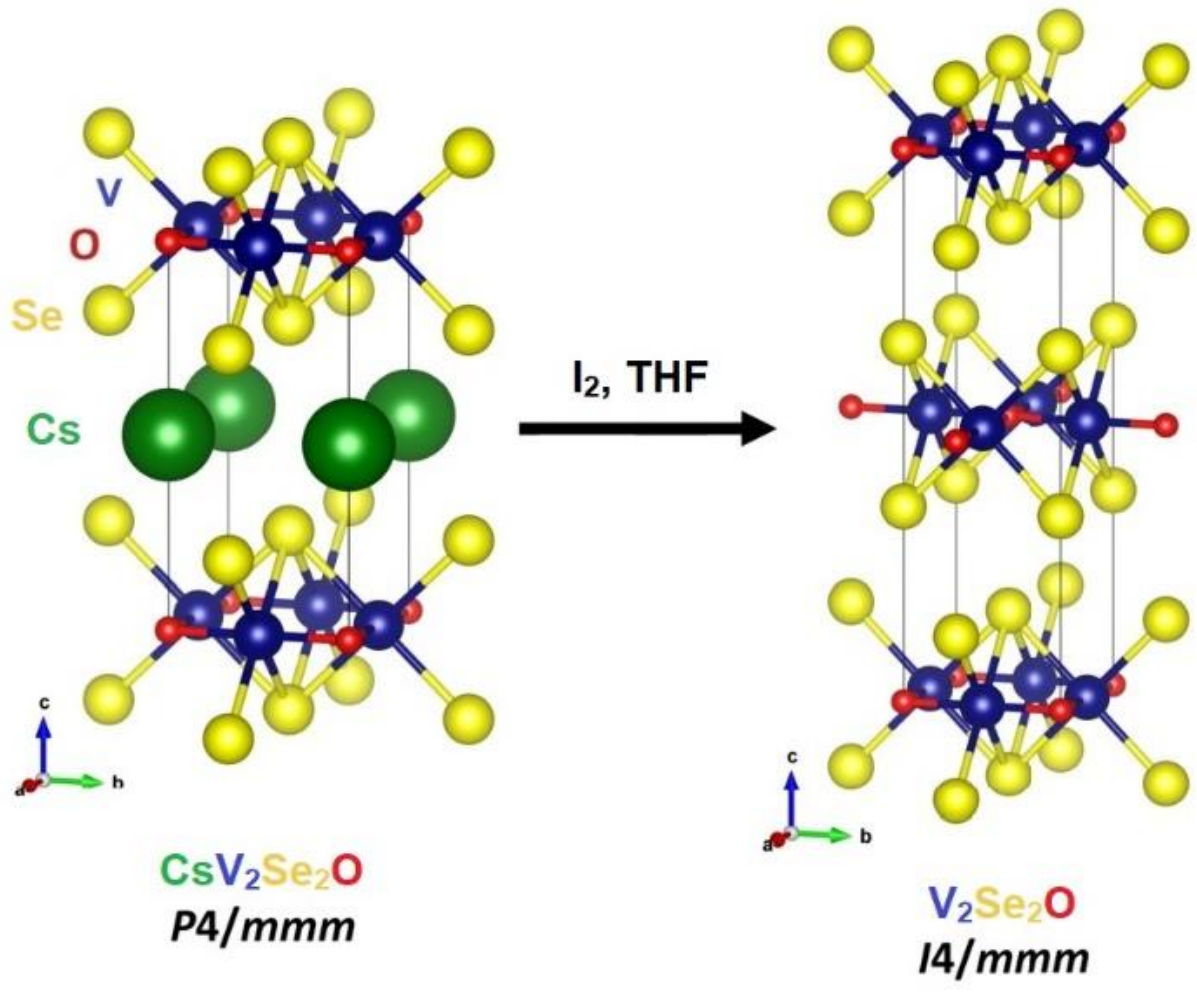

*Figure 8: Schematic showing the deintercalation of $Cs^+$ ions from $CsV_2Se_2O$ to produce the van der Waals phase $V_2Se_2O$.*

The work of Lin *et al*. was followed in 2018 by Ablimit *et al*. who found that $Rb^+$ ions could be extracted from $RbV_2Te_2O$ using milder conditions of $H_2O$ at room temperature [71]:

$$RbV_2Te_2O + H_2O \longrightarrow V_2Te_2O + RbOH + \frac{1}{2}H_2$$

The product, $V_2Te_2O$, is a van der Waals phase isostructural with $V_2Se_2O$ but is metallic, with a $T^2$ dependence of the resistivity at low temperatures. Like $V_2Se_2O$, it is air-stable, likely because +3 is a favoured oxidation state of vanadium ions. Unfortunately, neither charge-density-wave behaviour nor superconductivity has been observed down to 30 mK. $V_2Te_2O$ can also be produced from $KV_2Te_2O + H_2O$ in the same manner [24].

Finally, the isostructural $Ti_2Te_2O$ can be produced by treating either $KTi_2Te_2O$ or $RbTi_2Te_2O$ with water. It has a structural and electronic transition at approximately 250 K, where the magnetic susceptibility drops on cooling before becoming constant again, but further physical-property investigations have been hampered by its highly air-sensitive nature [24], in contrast with $V_2Se_2O$ and $V_2Te_2O$. The shape of the magnetic susceptibility anomaly for $Ti_2Te_2O$ strongly resembles that of $KV_2Se_2O$ [67], which suggests that similar electronic factors may be at play in directing the magnetic properties of these compounds, despite the different identity and electron count of the transition metal.

None of these three van der Waals compounds has been successfully synthesised using direct combination of elements and binary precursors at high temperatures. Their only synthesis is via low-temperature topotactic reactions [24,71]. This stands in stark contrast to the more well-known van der Waals chalcogenides, such as $Bi_2Se_3$, FeSe, $MoS_2$, etc. which are easily synthesised from the elements. The relative instability of these oxychalcogenides may arise from their unique mixed-anion chemistry, in that a mixture of binary chalcogenides and binary oxides is thermodynamically more stable than the mixed-anion target phase. Evidence for this hypothesis is provided by the decomposition of $V_2Te_2O$ under vacuum, and $Ti_2Te_2O$ under argon, at relatively low temperatures (400-500 °C) [24,71].

### *4.3 Intercalation reactions of $V_2Te_2O$.*

Reductive intercalation of layered van der Waals materials is a versatile method for altering their physical properties. For example, the Curie temperature of $CrGeTe_3$ rises from 66 K to 240 K upon intercalation of sodium ions [72] and the superconducting transition temperature of FeSe rises from 9 K to 43 K upon intercalation of Li ions in ammonia [73]. Such reactions are typically carried out at low temperatures so that the reaction proceeds topochemically, i.e. without destroying the structure of each layer, but simply inserting the intercalant species between the layers. This can be done using a range of chemical reagents, including but not limited to vapourised metal ions, solvated metal ions, and organometallic compounds. For further discussion of this area, the reader is directed to a recent review by Witte *et al.* on tuning magnetism in van der Waals materials using intercalation reactions [74].

Vanadium oxytelluride $V_2Te_2O$, itself formed by de-intercalation of alkali metals from the parent (K/Rb)$V_2Te_2O$ (Section 4.2), can undergo intercalation reactions with K, Rb or Cs metal dissolved in liquid ammonia [75]. These reactions re-form the primitive parent material, albeit with significant reflection-dependent peak broadening in the powder X-ray diffraction patterns, reflecting compositional inhomogeneity [76]. This could be alleviated upon low-temperature annealing to give alkali-deficient products $K_xV_2Te_2O$ ($x \sim 0.6$). Attractive though this proof-of-concept "reversibility" may appear, however, partially alkali-deficient products can already be synthesised directly using the ceramic process with appropriate molar amounts of the alkali metal reagent [24]. Both routes, the direct synthesis and intercalation followed by annealing, suffer from difficulty in controlling the final value of *x*.

Unlike their analogues with larger alkali metals, "$LiV_2Te_2O$" and "$NaV_2Te_2O$" could not be synthesised at high temperatures, probably because Li and Na are too small to occupy the 8-coordinate *A* site in the $AM_2Q_2O$ structure [75]. However, lithium ions can be intercalated into $V_2Te_2O$ both chemically, using n-butyllithium solution, and electrochemically [77]. The crystal structure of the product, $Li_xV_2Te_2O$, was solved from powder neutron diffraction data, revealing the stoichiometry ($x = 0.94(2)$) and hence the average oxidation state of vanadium, +2.53. Lattice parameters obtained from additional powder X-ray diffraction suggested that further reduction of vanadium should be possible via both synthetic routes, Figure 9(a). The crystal structure of $Li_xV_2Te_2O$ is very similar to the parent $V_2Te_2O$, with the same space group (*I*4/*mmm*) and an expanded unit cell; the intercalated $Li^+$ ions occupy six-coordinate sites in the van der Waals gaps, Figure 9(b).

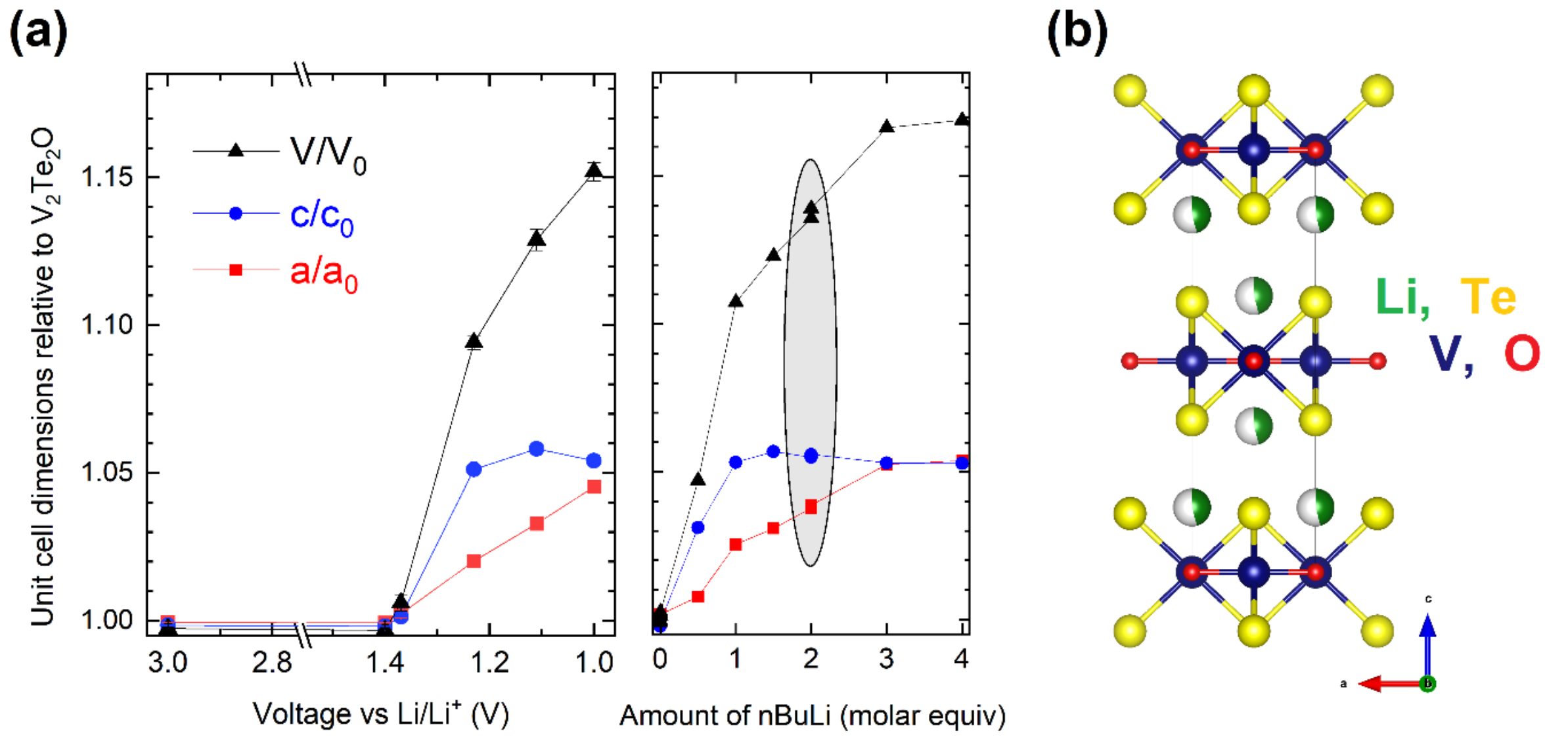

*Figure 9: (a) Lattice parameters of $Li_xV_2Te_2O$ samples made electrochemically and chemically. The lattice parameters for the sample investigated by neutron diffraction are highlighted in grey. (b) Crystal structure of $Li_{0.94(2)}V_2Te_2O$ determined from powder neutron diffraction. Adapted from reference* [77]. *The half-coloured spheres for Li indicate the fractional atomic occupancy of those sites, which are on average 47% occupied by Li and 53% vacancies.*

The smooth variation in lattice parameters upon intercalation of $Li^+$ indicates that solid solutions $Li_xV_2Te_2O$ should be formed for a range of *x* values in both the chemical and electrochemical methods. Thus, the oxidation state of vanadium, and therefore its d-electron count, may be smoothly tuned from +3 down to close to +2. This technique is therefore a useful tool in the search for superconducting oxychalcogenide analogues of the titanium oxypnictides. In the oxypnictides, besides physical pressure, substitution of $Ba^{2+}$ has been used to exert chemical pressure, with the aim of tuning the transition metal electron count and hence $T_C$ [8–11]. However, this has only been done during the initial synthesis stage. In the oxychalcogenides, with their more labile alkali metal ions, topochemical reactions and electrochemistry offer interesting ways to tune the metal electron count, and are worthy of future study. Furthermore, topochemical ion-exchange reactions should be explored as an alternative to direct reactions of the alkali metal, in order to synthesise Li- and Na-containing materials in fewer steps [78].

### *4.4 Future research directions in low-dimensional materials.*

Monolayers of the anti-cuprate type van der Waals systems, including $Ti_2Te_2O$, have recently been predicted to host unusual quantum Hall states depending on the type of magnetic order [26,79]. Furthermore, monolayer $V_2Te_2O$, $V_2S_2O$, and the mixed-chalcogenide or "Janus" monolayers $V_2STeO$, $V_2SSeO$ and $V_2SeTeO$ (Figure 10), have been proposed as candidate altermagnetic materials with coexisting piezoelectricity and piezomagnetism [80,81]. Zou *et al.* have also calculated the effects of substitution of later transition metals, Cr, Mn, Fe, Co, Ni, into the monolayer $M_2Q_2O$ systems [82]. They found that all $M_2Q_2O$ phases ($M$ = Ti–Ni) are dynamically stable apart from $Co_2Te_2O$, and all are potential candidates for altermagnetism because of the two magnetic sublattices which are inequivalent under translational symmetry. Six compositions exhibited a ferroelectric distortion from an off-centre oxide anion, which is reminiscent of the structural distortion observed in bulk $Ln_2O_2Mn_2Se_2O$ [34][82]. Of the stable monolayers, the cleavage energies (particularly with earlier transition metals: Ti, V and especially Cr) were found to be comparable with the cleavage energies of previously studied materials such as graphene and $MoS_2$, which suggests that mechanical exfoliation of anti-cuprate materials may be used to develop novel low-dimensional devices.

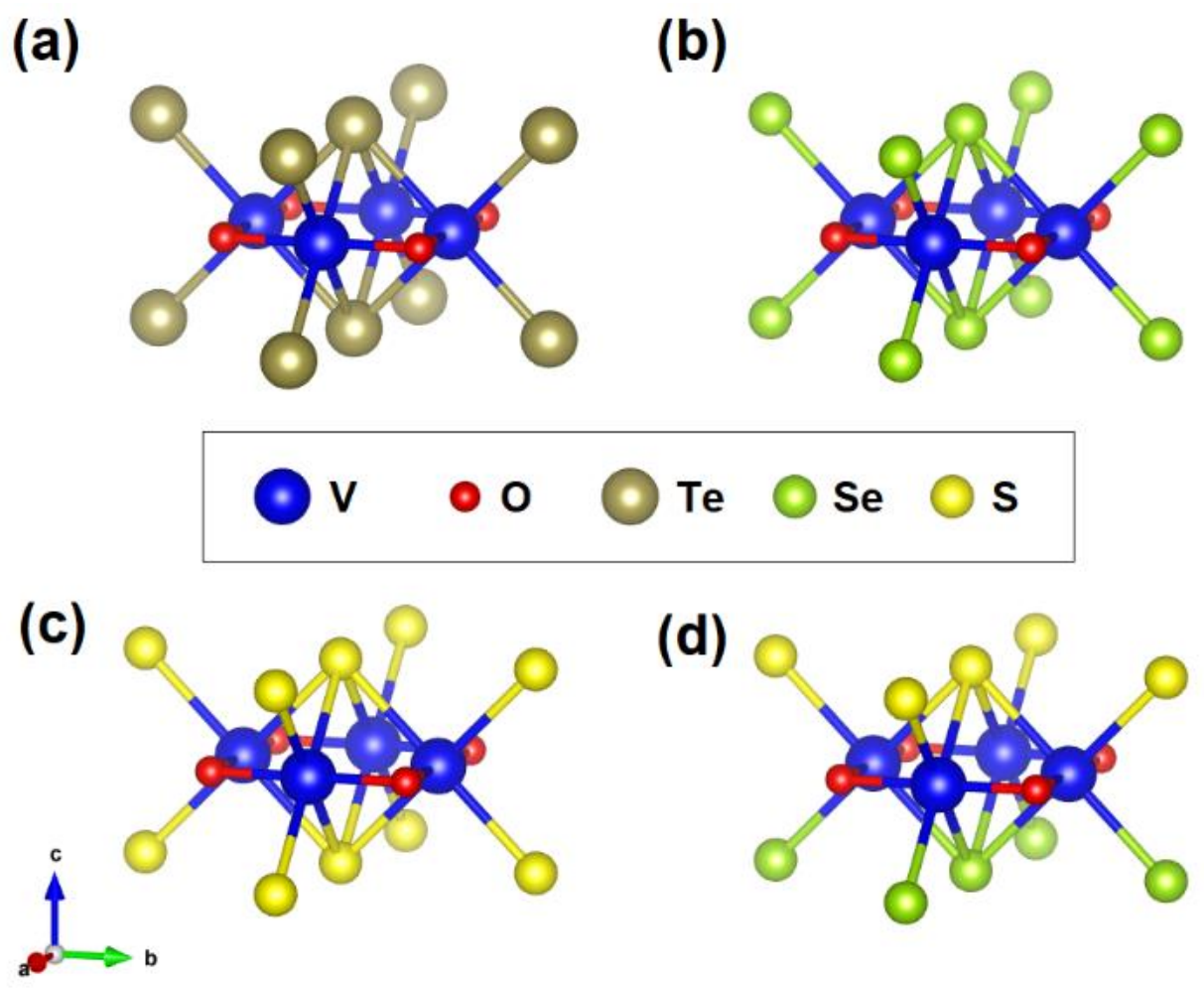

*Figure 10: Unit cells of the predicted monolayer materials (a) $V_2Te_2O$, (b) $V_2Se_2O$, (c) $V_2S_2O$, (d) $V_2SSeO$.*

In conclusion, these van der Waals materials offer an exciting opportunity for novel experimental physics if they can be either deposited or exfoliated from the bulk into monolayers. Chemical exfoliation from the alkali-metal-containing parent materials, as used for the vanadium oxychloride VOCl [83], is another promising strategy which should be explored for these layered oxychalcogenides.

## 5 Outlier materials and the landscape for future research

The compositions described in the previous sections encompass the transition metal ions $M$ = Ti, V, Mn, Fe, and Co. At the left-hand side of the $d$-block, where the metal ions tend to be in the 3+ oxidation state, the [$Ti_2O$] anti-cuprate square lattice has been extensively explored in oxypnictide materials [20] as well as in the metallic oxychalcogenides discussed here. Further along the series, oxychalcogenide materials have been discovered and characterised that contain $Mn^{2+}$, $Fe^{2+}$ or $Co^{2+}$ in the anti-cuprate layer, with several options for charge-balancing "spacer" layers with different structures and chemical bonding.

There is a clear gap between V and Mn. Whilst partial Cr substitution has been successful in forming $BaTi_{2-x}Cr_xAs_2O$ (but only up to a maximum $x = 0.154$) [7], the chromium oxypnictides $Sr_2CrO_2Cr_2As_2O$ [21] and $La_2Cr_2As_2O_yH_x$ [22] remain the only examples so far of full Cr substitution on the anti-cuprate square lattice; no oxychalcogenides have been reported with a [$Cr_2O$] layer. In the solid state as well as aqueous solutions, $Cr^{2+}$ is strongly reducing and $Cr^{3+}$ is by far more stable [84], which may explain the current absence of oxychalcogenides containing [$Cr_2O$] layers. However, this raises the interesting possibility of Cr(III) analogues of the V(III) and Ti(III) van der Waals compounds. Even if such target materials cannot be made topochemically from less-oxidised parent materials, monolayers might be achievable using atomic deposition techniques. The stability of these phases, i.e. $Cr_2S_2O$, $Cr_2Se_2O$ and $Cr_2Te_2O$, has been computed [82] and suggests that they are appropriate synthetic targets. Furthermore, the schematic relationship between the [$Cr_2As_2$] and [$Cr_2As_2O$] layers – viewed as the insertion of an oxygen anion and the formal oxidation of Cr, as discussed by Sheath *et al.* [21] and Park *et al.* [22] – provides a very interesting perspective for the future design of new compounds containing an anti-cuprate type layer, perhaps utilising high-pressure apparatus.

Considering the large number of Mn, Fe and Co oxychalcogenides (Section 2), the absence of $Ni^{2+}$ analogues is somewhat surprising. However, Le *et al.* have recently put forward compounds containing the $[Ni_2Q_2O]^{2-}$ anti-cuprate layer as potential high temperature superconductors under chemical or physical doping regimes [85]. The $3d^8$ electronic configuration of $Ni^{2+}$ corresponds to a $(t_{2g})^6(e_g)^2$ arrangement in an octahedral crystal field. Unconventional high-temperature superconductivity like that of the cuprates is said to require isolated *d*-orbitals, close to the Fermi energy, which display strong in-plane coupling to the neighbouring anion *p*-orbitals. This is not the case for perovskite-like [$NiO_2$] layers, but it is proposed to be achievable for the anti-cuprate layer structure, with *trans*-$NiQ_4O_2$ coordination environments with the octahedral axis aligned in-plane rather than out-of-plane. DFT calculations carried out on the composition $La_2O_2Ni_2Se_2O$ predicted it to be a stable compound with a C-type striped collinear spin structure. This magnetic structure is dominated by the AFM $J_2$' interactions (180° Ni–O–Ni), which are much larger than those calculated for the Fe and Co analogues, hence the different magnetic structures favoured (Fe compounds: 2-**k** model, Co compounds: typically G-type AFM). Considering these computational results, $La_2O_2Ni_2Se_2O$, its S or Te analogues, and related compositions such as $Ba_2F_2Ni_2Se_2O$ are the obvious next targets for solid-state chemists wishing to synthesise new anti-cuprate materials. Moving into the 2D regime, $Ni_2Q_2O$ monolayers are predicted to be stable, albeit with higher cleavage energies than the corresponding materials with other transition metals [82].

Beyond nickel, we can finally consider copper. Whilst the cuprates are numerous and familiar, anti-cuprates containing [$Cu_2O$] layers are limited to only one material. In 1993, Park *et al.* began their report: "*Solid-state oxy-chalcogenides of the late transition metals are rare. One reason for this may be the relative low oxophilicity and high chalcophilicity of these metals*" [86]. In the subsequent decades, some new families of copper oxychalcogenides have been reported including $(Sr/Ba)_2MO_2Cu_2Q_2$ ($M$ = Mn, Co, Ni; $Q$ = S, Se, Te) [87–89] and $(Sr/Ba)_3Sc_2O_5Cu_2Se_2$ [90,91]; however, none of those contains an anti-cuprate layer like the one found in the first reported example, $Na_{1.9}Cu_2Se_2Cu_2O$. This compound was first isolated from a Cu/$Na_2Se_x$ flux with trace oxygen impurities. Its crystal structure (Figure 11) is described as an intergrowth of "*independent but isoperiodic*" [86] layers of $[Cu_2Se_2]^{2-}$ and neutral [$Cu_2O$], separated by single layers of $Na^+$ ions. As such, it differs from all of the other compounds considered in this review, which have the [$M_2O$] square layer directly sandwiched by chalcogenide ions. The observed metallic conductivity of $Na_{1.9}Cu_2Se_2Cu_2O$ was rationalised by its slight Na deficiency, which causes partial oxidation of copper ions (holes in the valence band); calculations of the density of states suggested that these holes are localised in the more electron-rich [$Cu_2Se_2$] layers [92].

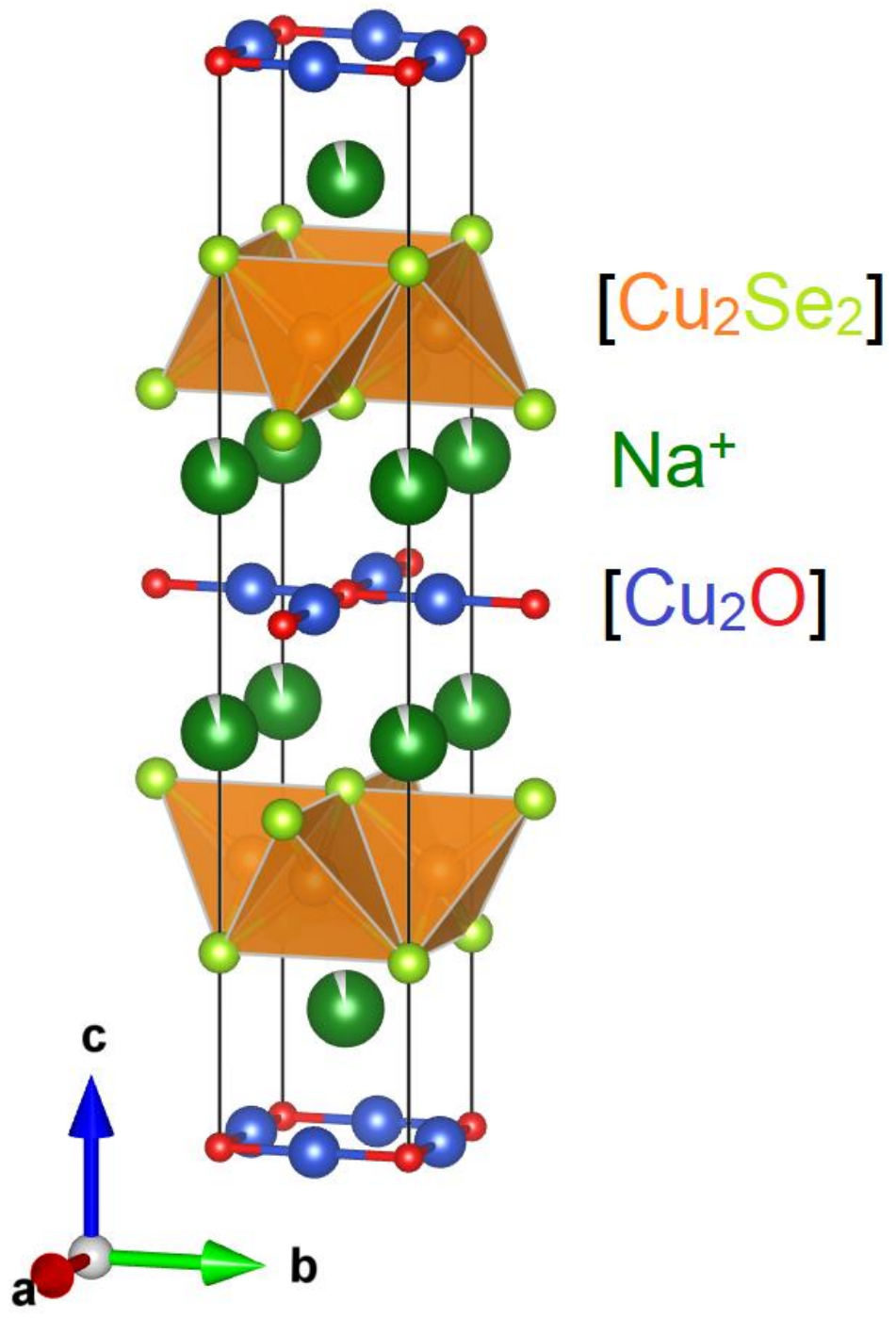

*Figure 11: Crystal structure of $Na_{1.9}Cu_2Se_2Cu_2O$ (space group I4/mmm) showing the anti-cuprate [$Cu_2O$] layers and the anti-PbO-type [$Cu_2Se_2$] layers consisting of edge-sharing $CuSe_4$ tetrahedra. The partially-coloured spheres for Na indicate the fractional atomic occupancy of those sites (95%)* [86].

The existence of $Na_{1.9}Cu_2Se_2Cu_2O$ demonstrates the potential for inclusion of another "building block" [19] in the design of new anti-cuprate materials, namely the [$Cu_2Se_2$] layer, which has already been explored in materials such as $(Sr/Ba)_2MO_2Cu_2Se_2$ and $(Sr/Ba)_3Sc_2O_5Cu_2Se_2$. It has the anti-PbO structure type with tetrahedral Cu ions and square-pyramidal Se ions [86]. In their original paper, Park *et al.* pointed out the "fortuitous" correspondence between the in-plane lattice parameters of the [$Cu_2Se$] and [$Cu_2O$] blocks, which may not be possible to replicate with other transition metals, particularly those early in the 3*d* series. However, new copper oxychalcogenides based on the parent $Na_{1.9}Cu_2Se_2Cu_2O$ structure could be discovered by utilising soft chemistry methods for partial de-intercalation of Cu [88,93] or even fully "collapsing" the $[Cu_2Se_2]$ layer into perselenide dimers [94], which may then permit re-intercalation of other alkali metal species [95]. With the assistance of high-pressure synthesis, Ag ions might also be substituted for Cu, and/or Te substituted for Se, in the $[Cu_2Se_2]$ block [96] with the aim of synthesising anti-cuprate compounds with a combination of 3*d* and 4*d* transition metal ions. The number of possible combinations of different "building block" layers is very large, particularly when soft-chemical methods are added to the synthetic toolbox, so computational methods should be used to help screen potential new compositions for stability.

## 6 Conclusions

Materials containing the anti-cuprate square lattice of transition metal and oxide ions are a diverse family of compounds with great variety in their structures, chemistry, and physical properties. Variation of the transition metal across the 3*d* series permits the $[M_2Q_2O]$ layer to take the total charge of –2 or –1, which in turn allows for a considerable number of compounds with different intervening "spacer" layers, as long as the total charge is neutral. These materials are structurally related to both the cuprate and titanium oxypnictide superconductors, and their own physical properties may be tuned by chemical substitution on the *M*, *Q*, or other atomic sites in the crystal structure, allowing the metal ions to take various oxidation states between +2 and +3, including fractional oxidation states. Furthermore, soft chemistry has been shown to be a powerful tool for oxidation and reduction reactions of these materials, producing novel metastable systems with unusual properties.

Overall, this series of materials offers a versatile playground to explore the interplay between structure and electronic and magnetic properties in layered solids. Ion-exchange and molecular intercalation reactions are proposed as promising future research avenues for discovering new materials in this family, and recent theoretical studies have also highlighted their potential as altermagnetic materials, either in bulk crystals, or in monolayers if they can be deposited or chemically or physically exfoliated. Pressure may also be used for tuning physical properties in these materials. Given the interdisciplinary nature of this research area, collaboration and communication between materials chemists and condensed-matter physicists is vital to advancing the field.

## Data Availability Statement

No new data were generated for this article.

## Acknowledgements

I gratefully acknowledge research funding from Jesus College, Cambridge, and from the US National Science Foundation (NSF) Grant Number 2201516 under the Accelnet program of Office of International Science and Engineering (OISE).